\documentclass{sigchi}

\usepackage{balance}    
\usepackage{graphics}   
\usepackage{times}      
\usepackage{url}        
\usepackage{enumerate}
\usepackage{csquotes}
\usepackage{subfigure}
\usepackage{needspace}  
\usepackage{array}
\usepackage{multirow}
\usepackage{tabu}

\makeatletter
\def\url@leostyle{%
  \@ifundefined{selectfont}{\def\UrlFont{\sf}}{\def\UrlFont{\small\bf\ttfamily}}}
\makeatother
\def\pprw{8.5in}
\def\pprh{11in}

\newcommand\tabhead[1]{\small\textbf{#1}}
\newcolumntype{L}[1]{>{\raggedright\let\newline\\\arraybackslash\hspace{0pt}}m{#1}}
\newcolumntype{C}[1]{>{\centering\let\newline\\\arraybackslash\hspace{0pt}}m{#1}}
\newcolumntype{R}[1]{>{\raggedleft\let\newline\\\arraybackslash\hspace{0pt}}m{#1}}
\newcolumntype{H}{>{\setbox0=\hbox\bgroup}c<{\egroup}@{}}

\usepackage{pdfcomment}
\usepackage[rgb]{xcolor}

\usepackage[pdftex]{hyperref}
\hypersetup{
pdftitle={An Interactive Tool for Natural Language Processing on Clinical Text},
pdfauthor={Gaurav Trivedi},
pdfkeywords={Electronic medical records, visualization, interactive machine learning, text analysis, natural language processing},
bookmarksnumbered,
pdfstartview={FitH},
colorlinks,
citecolor=black,
filecolor=black,
linkcolor=black,
urlcolor=black,
breaklinks=true,
}

\newcommand{\hhcomment}[1]{}
\newcommand{\hhcommentRevisit}[1]{}
\newcommand{\hhcommentFixed}[1]{}

\newcommand{\rhcomment}[1]{}
\newcommand{\rhcommentRevisit}[1]{}
\newcommand{\rhcommentFixed}[1]{}

\newcommand{\gtcomment}[1]{}
\newcommand{\gtcommentRevisit}[1]{}
\newcommand{\gtcommentFixed}[1]{}

\begin{document}

\title{An Interactive Tool for Natural Language Processing on Clinical Text}

\numberofauthors{6}
\author{
  \alignauthor Gaurav Trivedi\\
    \affaddr{Intelligent Systems Program}\\
    \affaddr{University of Pittsburgh}\\
    \email{trivedigaurav@pitt.edu}
  \alignauthor Phuong Pham\\
    \affaddr{Dept. of Computer Science}\\
    \affaddr{University of Pittsburgh}\\
    \email{phuongpham@cs.pitt.edu}
  \alignauthor Wendy Chapman\\
    \affaddr{Dept. of Biomed. Informatics}\\
    \affaddr{University of Utah}\\
    \email{wendy.chapman@utah.edu}
  \alignauthor Rebecca Hwa\\
    \affaddr{Dept. of Computer Science \\  \& Intelligent Systems Program}\\
    \affaddr{University of Pittsburgh}\\
    \email{hwa@cs.pitt.edu}
  \alignauthor Janyce Wiebe\\
    \affaddr{Dept. of Computer Science \\ \& Intelligent Systems Program}\\
    \affaddr{University of Pittsburgh}\\
    \email{wiebe@cs.pitt.edu}
  \alignauthor Harry Hochheiser\\
    \affaddr{Dept. of Biomed. Informatics \\ \& Intelligent Systems Program}\\
    \affaddr{University of Pittsburgh}\\
    \email{harryh@pitt.edu}
}
\maketitle

\begin{abstract}
Natural Language Processing (NLP) systems often make use of machine learning techniques that are unfamiliar to end-users who are interested in analyzing clinical records. 
Although NLP has been widely used in extracting information from clinical text, current systems generally do not 
support model revision based on feedback from domain experts. 

We present a prototype tool that allows end users to visualize and review the outputs of an NLP system that extracts binary variables from clinical text. 
Our tool combines multiple visualizations to help the users understand these results and make any necessary corrections, thus forming a feedback loop and helping improve the accuracy of the NLP models. We have tested our prototype in a formative think-aloud user study with clinicians and researchers involved in colonoscopy research. Results from semi-structured interviews and a System Usability Scale (SUS) analysis show that the users are able to quickly start refining NLP models, despite having very little or no experience with machine learning. Observations from these sessions suggest revisions to the interface to better support review workflow and interpretation of results. 
\end{abstract}


\category{H.5.2.}{Graphical User Interface}{User Interfaces}\category{H.5.m.}{Information Interfaces and Presentation (e.g. HCI)}{Miscellaneous}.

\gtcomment{I will replace this figure with one that has annotations marking these views - The captions needs to be revised as well}
\begin{figure*}[!ht]
\centering
\includegraphics[width=2.1\columnwidth]{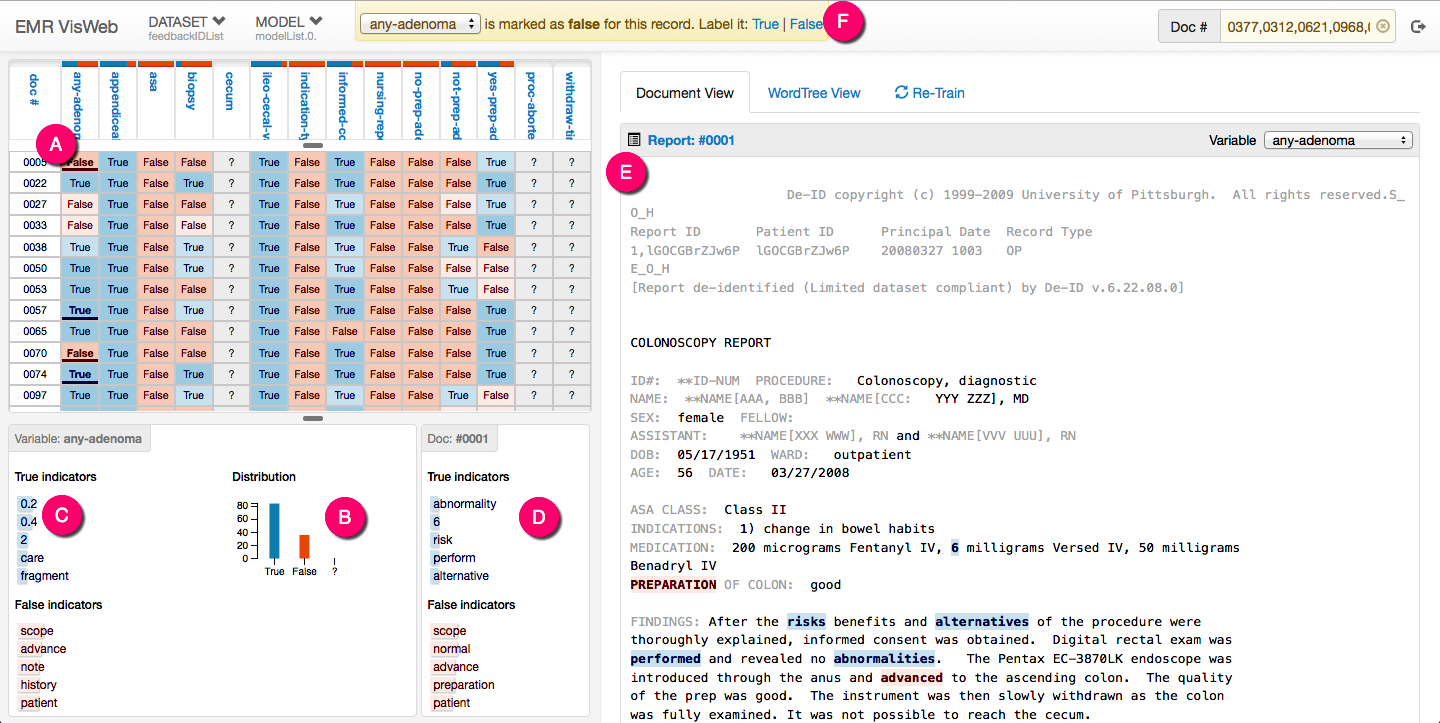}
\caption{ (a) The \textit{grid view} shows the extracted variables in columns and individual documents in rows, providing an overview of NLP results. Below the grid, we have statistics about the active variable with (b) the distribution of the classifications and (c) the list of top indicators aggregated across all the documents in the dataset. (d) Indicators from the active report are shown on the right. (e) The \textit{document view} shows the full-text of the patient reports with the indicator terms highlighted. (f) Feedback can be sent using the yellow control bar on the top, or by using the right-click context menu.}
\label{fig:main}
\end{figure*}

\needspace{2em}\section{Introduction and Background}
Electronic Health Records (EHRs) are organized collections of information about individual patients. They are designed such that they can be shared across different settings for providing health care services. The Institute of Medicine committee on improving the patient record has recognized the importance of using EHRs to inform \emph{decision support systems} and support data-driven quality measures \cite{oldbook}. One of the biggest challenges in achieving this goal is the difficulty of extracting information from large quantities of EHR data stored as unstructured free text. Clinicians often make use of narratives and first-person stories to document interactions, findings and analyses in patient cases \cite{narratives}. As a result, finding information from these volumes of health care records typically requires the use of NLP techniques 
to automate the extraction process.

There has been a long history of research in the application of NLP methods in the clinical domain \cite{barriers}. Researchers have developed models for automatically detecting outbreak of diseases such as influenza \cite{yeye}, identifying adverse drug reactions \cite{adr-iyer, adr-ramirez, adr-medlee}, and measuring the quality of colonoscopy procedures \cite{harkema}, among others. Due to the complexity of clinical text, the accuracy of these techniques may vary \cite{nlpintro}. Current tools also lack provision for end-users to inspect NLP outcomes and make corrections that might improve these results. Due to these factors, Chapman et. al. \cite{barriers} have identified ``lack of user-centered development" as one of the barriers in NLP adoption in the clinical domain. There is a need to focus on development of NLP systems that are not only generalizable for use in different tasks but are also usable without excessive dependence on NLP developers. In this paper, we have explored the design of user-interfaces for use by end users (clinicians and clinical researchers) to support the review and annotation of clinical text using natural language processing. 


We have developed an interactive web-based tool that facilitates both the review of binary variables extracted from clinical records, and the provision of feedback that can be used to improve the accuracy of NLP models. Our goal is to close the natural language processing gap by providing clinical researchers with highly-usable tools that will facilitate the process of reviewing NLP output, identifying errors in model prediction, and providing feedback that can be used to retrain or extend models to make them more effective. 


\needspace{2em}\section{Related Work} 

During the process of developing our interactive text analysis tool for clinical domain, we studied relevant work from multiple research areas spanning across different domains, such as \emph{Visualization}, \emph{Interactive Machine Learning} and \emph{Interface Design}. We have built upon the following work in these areas for the design of our tool.

\needspace{2em}\subsection{Visualization and Sensemaking}
Visualization tools such as WordTree \cite{WordTree} and Tiara \cite{tiara}\hhcommentFixed{better to give the names up front} help in providing a visual summary of large amount of text data. While Tiara focuses on content evolution of each topic over time, WordTree provides a keyword in context method of exploring the text. Other tools such as Jigsaw \cite{jigsaw} help users interpret document collections by visualizing documents in multiple graph, cluster and list views.\hhcommentFixed{say something about the goals of Jigsaw. What you could really say would be that Jigsaw focus on helping users build integrative understanding of many documents in a collection.} Our task in reviewing clinical documents is somewhat different, in that our goals are to understand common textual patterns and to use those patterns to improve NLP models. We have adapted elements of these views - in particular, WordTree's phrase view and Jigsaw's document view to support our goals.  \hhcommentFixed{I have rephrased the sentence to say "we have built on these ideas" -- have we really incorporated many? Be specific-- or just remove this sentence - possibly better -- We have indeed built on these ideas --> WordTree and Jigsaw's document view are most evident.} The purpose of these visualizations would be to provide both detailed document-level views and also dataset-level overviews. \hhcommentFixed{which visualizations? Jigsaw? Consider removing this sentencedisregard the rest of this comment...I like this piece, but I would end this paragraph with it. Introduce the other tools by saying how many tools for visualizing text have been developed. Can you cite some other papers?}

\needspace{2em}\subsection{Interactive Machine Learning}\hhcommentFixed{a lot is a bit colloquial. Say there has been a significant body of work, or many efforts have developed...}There have been many efforts to develop user-centric tools for machine learning and NLP making it easier for the end users to build models. \hhcommentFixed{a bit awkward, but ok} D'Avolio et. al. \cite{avolio} have described a prototype that combines several existing tools such as Knowtator \cite{knowtator} for creating text annotations, and cTAKES \cite{ctakes} for deriving NLP features, within a common user interface that can be used to configure the machine learning algorithms and export their results. Our present work complements this effort, focusing instead on facilitating expert review of NLP results and provision of feedback regarding the accuracy and completeness of details extracted from NLP data. 

Other efforts have taken this idea even further to build interactive machine learning systems that learn iteratively from their end-users. Sometimes referred to as \emph{``human-in-the-loop''} methods, these techniques involve a learning system whose output is used by the end-user to further inform the system about the learning task. This forms a closed loop that can be used to build continuously improving models of prediction. Some examples include applications in interactive document clustering \cite{iml-apolo}, document retrieval \cite{iml-heimerl}, image segmentation \cite{iml-image}, bug triaging \cite{iml-cuet} and even music composition \cite{iml-music}. These successes suggest that it may be promising to use feedback to improve machine learning models in the clinical domain.\hhcommentFixed{good!}

\needspace{2em}\section{Design Requirements}
\rhcomment{to the uninitiated, the NLP/ML relationship/connection isn't clear. in other words, why does it make a difference whether the user knows anything about ML or not? In general, my feeling about this paragraph is that most of the points it tries to make are relevant, but somehow the paragraph as a whole seems to not unite. I like the summary bullet points. I'd like to see the paragraph somehow preparing the reader to understand/arrive at the same summary points at the end. Right now, this paragraph sort of just says the same thing as the bullets in paragraph form. See my suggested outline.}We assume that the users of our tool are domain experts who are familiar with the contents of the documents being reviewed, but not with machine learning. \hhcommentFixed{a bit awkward} Our approach focuses on designing interaction methods and novel data visualizations for the user to interact with and correct the learning models. \hhcommentFixed{why? - it's fine now} Further, while most of the focus in previous work has been towards developing usable interaction methods with the learning algorithms, more often in real world applications, we find that obtaining reliable labels for the training examples is very difficult, costly or time-consuming. In domains such as medicine, we require the help of skilled domain experts. Labeled data are important to support training automated systems; yet, large amounts of training data do not exist for new use cases or for applications that may arise in the future. It is therefore of great practical interest to develop methods for obtaining good quality labels efficiently. Such methods are even more in need for NLP applications because it is time consuming for annotators to obtain the  contextual information from the text before labeling. \hhcommentFixed{Is it okay now? I have incorporated Dr. Wiebe's suggestions -- all good, but should be explaind a bit more clearly in a separate paragraph. describe some of these issues a bit more clearly and systematically. Why are you bringing up these questions=== I think this is fine.} Lastly, we need to design techniques for the users to review the output of the NLP models. They should allow the users to find errors in predictions and make changes to build revised models. This would form a closed loop that would allow the users to iteratively create more accurate models that can be useful in their analysis. \hhcommentFixed{I have said that this would in building more accurate models iteratively; Does that make it clear? -- why? - make it clear that we need to do this to improve the models - yes. it's ok} These requirements are summarized as follows:

\begin{enumerate}[{R}1{:}]
\item The tool should facilitate interactive review of clinical text and make it easier for machine learning non-experts to work with NLP models. 

\item Visual presentation and interactive feedback components should facilitate the building of accurate NLP models.\hhcommentFixed{quicker than what? You might just fay "facilitate.} This includes providing efficient techniques for annotation and labeling the training examples and also for providing feedback that is consistent and informative. 

\item The interactive components should support the entire interactive machine learning loop - i.e. a \emph{review}, \emph{feedback} and \emph{retrain} cycle that can be used to revise NLP models iteratively.
\end{enumerate}


\needspace{2em}\section{Interface Design}


To demonstrate our tool, we have used an example dataset of colonoscopy reports by building on work done by Harkema et. al. \cite{harkema}. They have described an NLP system to extract values against a set of boolean variables from these reports. We have included a subset of 14 of these variables for the demo. Each patient record in the example dataset can include multiple linked reports from endoscopy and pathology. We have considered such reports together as a single document for learning and making predictions.

Figure~\ref{fig:main} shows a screenshot of our web-based tool. We have also uploaded a demo video of the tool at ~\url{http://vimeo.com/trivedigaurav/emr-demo}. In the following sections, we describe the individual components of the tool's user-interface, relating to the three requirements discussed above. \hhcommentFixed{We have descriptions of individual views and their use-cases - wouldn't that be sufficient? ---- can we have a pargraph describing how it will be used? add a scenario if space can be found.   - this paragraph is puzzling and hard to read. you could remove it and just go to the first subsection.  }

\needspace{2em}\subsection{Review}
An interactive machine learning cycle begins with the review step where the output from the learning model is shown to the user. Initial models can be trained on a few hand-annotated training examples. We have designed the following views in our tool to help the user inspect the prediction results.

\needspace{2em}\subsubsection{Grid View}
The grid-view is a table with columns showing the 14 variables and rows representing the individual documents. Each cell in the table shows the predicted value -- true or false -- corresponding to the particular  document-variable pair. This table is scrollable to accommodate all the documents in the dataset and extends beyond what is visible in Figure~\ref{fig:main}(a). We also have some cells with a question mark (?), where the model is unsure about the classification. This might happen for one of two reasons: either the classification algorithm does not identify a clear answer,  or the learning system does not have sufficient examples in the training data to make any predictions as yet. Subsequent feedback may tilt the classification in either direction. 

If the user hovers the mouse over a particular cell, a pop-up appears below it that shows the prediction probability, or how confident the learning system is in making that prediction. For example, the probability of a particular cell being true may be 75 percent. The grid also doubles up as a way to navigate through the documents. When a user clicks on a particular cell, the corresponding document-variable pair is activated in the all other views. The document view opens up the active document on the right-half of the screen. The highlighted cell in the grid indicates the currently active document-variable pair. Whenever the user clicks on a cell, we also mark it as visited to keep track of them. Visited cells in the grid are denoted by an asterisk symbol (*). \hhcommentFixed{why} 

An overview bar at the top of each column displays the true-false distribution (skew) of each variable. Exact distribution percentages are shown when the user mouses over the variable name.\hhcommentRevisit{I'll try to do the figures tomorrow -- can you emphasize this point in the figure? Perhaps with a numbered arrow?}

We have followed a uniform color scheme throughout the tool. Everything shaded blue represents a true value while the orange shades stand for false values. The colors were selected from a colorblind-safe palette.\hhcommentFixed{ be more specific - the coloring applies only to the cells and the words. Also, don't say truthy - just say true or false} In the grid view, the cells with higher probability have a darker background color. For example, a light blue cell indicates a low probability about a true classification, and a darker blue for a higher probability. 

\needspace{2em}\subsubsection{Views for displaying Keywords and Statistics}
Below the grid, we have views showing statistics about the currently active document and the variable. We show a histogram with a distribution of the true, false and unknown values over all the documents in the grid for the activated variable. Again, to reveal the exact counts under each prediction class, a user may hover the mouse over chart. \hhcommentFixed{don't mention this until you get to filtering - I think this is probably ok.} This display is similar to the overview bar above the grid but is more detailed and changes dynamically when the user uses the search box or the WordTree view to filter the document collection.

Our NLP pipeline uses a bag of words feature-set and a support vector machine (SVM) learning model for every variable, but it can be extended for use with different kinds of models and complement other existing tools as well. It works by identifying more informative features from a document (top terms) to make predictions. Informative terms are highlighted when present in the  current document in the right half, with overall distributions presented on the left-hand side of the screen. Terms are color-coded to indicate their contribution towards assigning the value of true or false against a variable, using colors from the document-variable grid. A mouse over each top term will reveal the feature weights from the learning system. Note that the current implementation consists of only unigram features but the same idea can be extended to $n$-grams as well. \hhcommentFixed{-what does the weighing scheme mean? clarify.  bit confusing here. Say that the models use bag-of-words to classify variables, and that this leads to scores for a given variable across all documents, and for the specific document for that variable. Note that we might have to say something about how this will work for non bag-of-words tools.  This still needs some work. .what does "weighing scheme mean? Are these weights assigned to terms by the NLP system?}

\needspace{2em}\subsubsection{Document View}
On the right hand side of the tool, we show the full-text of the reports. \hhcommentFixed{I have now introduced this before -- you have not previously talked about the two document types. add some description of these factors above where you introduce Harkema's data. Not fixed. you should mention the possibility of two reports/patient when you first introduce this datast. } The linked documents from a patient record, such as endoscopy and pathology reports are listed on the top of this view with shortcuts to jump to any of them. The top terms, both true and false (as seen in the grid view), are highlighted in the document. The keyword lists document the last view can be used to navigate through the report as well. Clicking on a keyword from the list causes the document to view to scroll to and highlight the first appearance of the term in the current document. This can also be done from the top terms list for the variables. However, since the term list contains the aggregate of the terms from all the documents together, there is a chance that a particular term doesn't appear in the open document. When such a term is clicked the keyword will be animated with a brief jitter to indicate that it cannot be found. The highlighted top-terms in the document view follow the same color scheme for true and false indicators.

Clinical reports also contain boilerplate text, or portions that can be considered to be having no effect on the predictions. These include de-identification headers, footers, and report's template text. These portions are dimmed in gray to improve the readability of the reports.


\begin{figure}
    \begin{center}
    \subfigure[The \textit{WordTree view} provides the ability to search for and explore word sequence patterns found across the documents in the corpus, and to provide feedback that will be used to retrain NLP models. In this example, we built the tree by searching for the word \emph{`biopsy'} and then drilled down upon the node \emph{`hot'}. The word tree now contains all the sentences in the dataset with the phrase \emph{`hot biopsy'} in them. It allows the user to get an idea of all the scenarios in which \emph{hot biopsy} has been used. Hovering over different nodes in the tree will highlight specific paths in the tree the selected term.]{\includegraphics[width=0.5\textwidth]{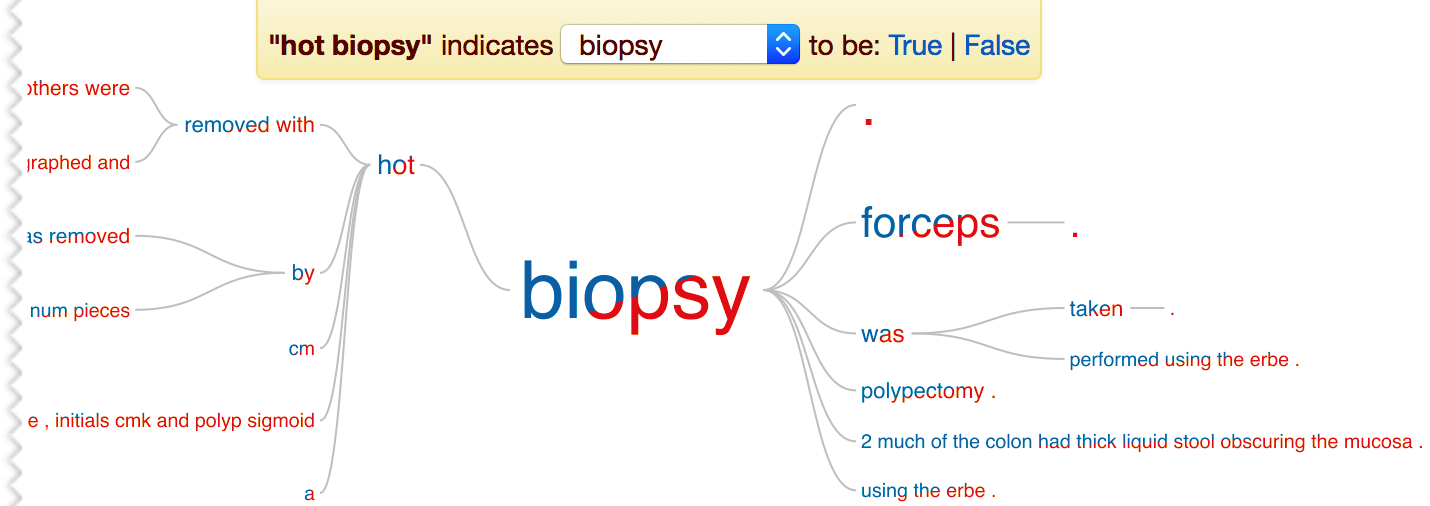}\label{fig:wordtree}}
    \subfigure[The \textit{Re-Train view} lists user-provided feedback, including any potential inconsistencies, and specifies changes in variable assignments due to retraining.]{\includegraphics[width=0.5\textwidth]{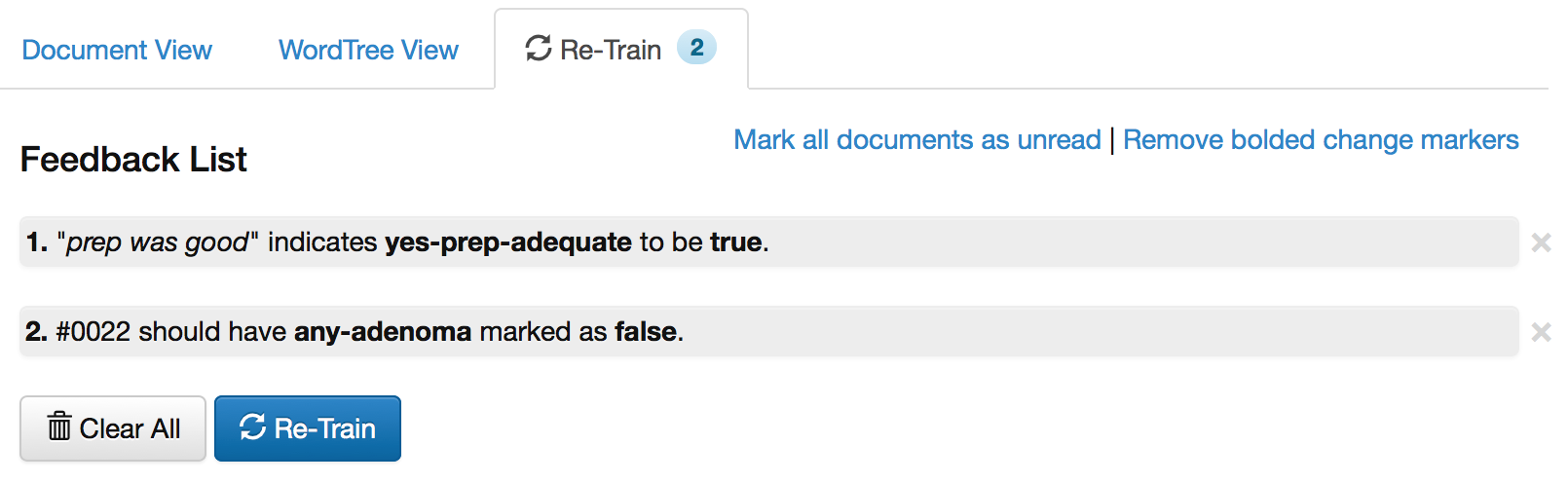}\label{fig:retrain}} 
    \caption{Screenshots of the \textit{WordTree} and \textit{Retrain} views.}
    \label{fig:other-views}
    \end{center}
\end{figure}

\needspace{2em}\subsubsection{WordTree View}
We have discussed views that provide detailed document-level visualization of the health records. But, we still need a visualization that could give a quick overview of the complete dataset. \hhcommentFixed{Please remove. It's not necessary. I think I'll retain this; Maintains a reading flow -- fine, but a bit informal. Consider removing} The WordTree \cite{WordTree} visualization offers a visual search tool for unstructured text that makes it easy to explore repetitive word phrases. The main advantage of using the wordtree is that it offers a complete data-set level visualization while retaining sentence level contextual information. A wordtree for a particular keyword consists of all the sentences in the dataset having that word or phrase. If one thinks of this keyword as the root of the tree, the branches represent the phrases that precede or follow that word. All the nodes of the tree are built recursively in this fashion.  The font size of a particular node is decided according to the total proportion of sentences that have it as a common starting phrase. 

We have made several improvements to the original wordtree design by Wattenberg and Viegas \cite{WordTree}. Their design restricts the root phrases to be present at the beginning or the end of a sentence. This allows the tree to grow only in one direction. We have used a modified design of the wordtree (as proposed in \cite{silverasm-github}) to construct a bi-directional tree that can grow in both directions. A sentence reads from left to right with the root phrase in the middle of the tree. Ends of sentences are denoted by a period (.) node. \hhcommentFixed{rephrased this. Does this read better? had been -A more of bluish color in the gradient would indicate a higher percentage of documents in the grid to be classified as true} This information is also conveyed in plain-text and numbers as a tool-tip on one corner of the WordTree view on mouse over. These gradients are dynamically updated according the variable selected by the user and also as the models change prediction upon retraining. The gradients provide an insight into how the machine learning model's prediction changes as different words or phrases are present or absent from the documents.

To start using the WordTree view, the user must enter a keyword or a phrase in the search bar. The tool creates a wordtree with the search query as the root after scanning all the sentences in the dataset. One can interact with the wordtree and navigate through its branches by clicking on individual nodes as shown in Figure~\ref{fig:wordtree}. Doing this prunes the tree and drill down into the details by adding the clicked node to be a part of the root node along with the search term. Clicking on the same node again reverts the view to the previous state of the tree. The gray bar below the wordtree, shows the number documents and and the percentage of the dataset represented in the tree. The WordTree view also has a full-screen mode which hides the other views in the tool when required.

We have extended the wordtree to use color-coded gradients to encode class distribution information. Each word is painted in a gradient, with the extent of the blue/orange color indicating the percentage of active documents in the grid classified as true/false, using the previously-described color palette. The grid view is also linked to the wordtree: pruning the branches in the tree  filters the set of documents to display in the grid to contain only those that are represented in the tree. The document ID list on the top right of the screen and the statistics views are similarly coordinated. If a user wishes to read more than what is available in the tree, they could just click on the corresponding cell in the grid to switch back to the document view and review the complete document. In the following section we will describe how the wordtree can be used for providing annotations as well.

\needspace{2em}\subsection{Feedback}
Feedback from the user is used by the system to improve upon the machine learning models by providing labels for documents that were previously not  part of the training set, or by correcting any misclassified documents. Useful feedback to the machine learning helps improve the prediction accuracy of the NLP models. \hhcommentFixed{move to beginning of paragraph} A user can provide feedback by simply changing the prediction class of a document for a given variable. \hhcommentFixed{for a given variable} The learning system would then be able to use this as a training example to learn from its features. The marginal benefits may be greater if a user classifies a group of documents together instead of annotating them one by one. To classify a group of documents, we could search for a selected text span and label all the matching documents as belonging to a certain class. \hhcommentFixed{good, but awkward probably ok} For this approach to work correctly, selected text spans  must convey consistent meanings across different usage scenarios in the dataset, and feedback based on these selections should not imply any contradictory classifications.   \hhcommentFixed{do we do all of those things? If not, what are we missing -- I don't think we are missing anything here.ok } 

Our prototype supports both multiple mechanisms for providing feedback and a review display that will alert the user to any potential inconsistencies associated with them.  To provide feedback for a particular document-variable pair, the user can select either true or false on the yellow control bar above the document view (Figure~\ref{fig:main}(f)). The currently active variable from the grid is pre-selected but the control also allows the user to quickly change the variable of interest by choosing from the drop-down options or by activating a different cell in the grid. \hhcommentFixed{or by clicking on a cell, right} 

Users can also provide more specific feedback by manually highlighting relevant text spans which could support document's classification. Like most other text annotators, the selection span automatically moves to ends of a word boundary if it is left hanging in the middle of a word. Multiple words forming a phrase can also be highlighted to be sent as a feedback.

Since the users are free to select their own text spans, there is a scope for the feedbacks to be inconsistent or prove to be less useful to the learning system. As a result, we have designed a feedback mechanism using the wordtree to provide them with some guidance in selecting these spans. Here the root phrase takes the role of the highlight span. Phrases are added before and after the root word as the user drills down the tree and prunes it. We believe that the wordtree may be useful while providing feedback step for it allows the users to give feedback on several documents together. The users are able explore the different use cases of a phrase in all the documents in the dataset with a single glance. It also helps in identifying tighter and more generic feedback phrases with its click to drill down design. If the user is able to make a choice without having to view the complete sentence, we can identify phrases that may be more important for the machine learning system. The varying font sizes of the phrases provide strong visual cues about their frequency of use in the dataset and thus encourages the users to attend to more useful phrases for training first. Further feedback for multiple documents based on a single phrase may help avoid potential conflict scenarios where the user may have highlighted similar keywords but selected different classes for feedback. To summarize, the wordtree not only provides an overview of the entire dataset but the provision of interacting with it allows the users to work directly with phrases and sentences in the dataset. It helps them to browse the data easily, send feedback actions to the learning systems, and see results with the help of color gradients.  \hhcommentFixed{Moved from the wordtree location to here -- I have rephrased it a bit; I wanted to discuss the fact that the views are not designed in isolation but work together  - Is it okay now? -- It's not necessary. I would remove it.-- consider removing paragraph or move description of utility of WT to introduction}\hhcommentFixed{consider trim}

All of three kinds of feedback can be submitted from the yellow bar present at the top of the screen that shows available options depending on the context. For example, an option for providing a feedback like this appears as soon as a text span is selected in the document. \hhcommentFixed{revise to say that users can do this by selecting text and right-clicking to a context menu} The document view also provide a right click menu as an additional affordance for the users to send feedback.

\needspace{2em}\subsection{Re-Train}
Re-training is the final step of the interactive machine learning cycle. The Re-Train view tab keeps a count of the number of feedbacks sent by the user and can be selected to view the list of proposed revisions to the model (Figure~\ref{fig:retrain}). The list includes all three kinds of feedbacks. Clicking on the re-train button launches the re-training process. Once the retraining is complete, a new model is created and the system updates the predictions in the grid and the linked views. The grid view indicates all the differences between the old and the new model predictions. One can spot these changes in bold. These cells will also have a bold underline in the bottom. This allows the user to identify changes made in the model as a result of their feedback.


The Re-Train view also provides guidance for resolving potentially contradictory feedback items. For example, a user may provide a particular text span indicating that a given document-variable pair should be set to be true, even as they label it false in another feedback setting. \hhcommentFixed{when can this happen - when must a document be labeled as true? also, we don't label documents -we label document/variable pairs} In these cases, the system will return an error message specifying the problem,  and highlight conflicting feedback items in red. These items can be revised or deleted from the Re-Train view, with red highlights disappearing when conflicts are resolved. Another conflict scenario involves the submission of suggested changes that undo the effects of earlier revisions to the model. These items are highlighted in yellow, and accompanied by an override option that will allow the newer input to take precedence over the earlier feedback. \hhcommentFixed{do you mean feedback that contradicts earlier feedback? Clarify}  The repeated re-training steps allow the users to build the learning models over several iterations. 

\needspace{2em}\section{Implementation and Deployment}
The system is implemented as a client-server architecture with communication over HTTP(S) using JSON. The user-interface has been built using the Angular (\url{angularjs.org}), \hhcommentFixed{cite the web site} D3 \cite{d3} and jQuery (\url{jquery.com}) javascript frameworks and libraries. The NLP learning system manages the model building and is deployed as a Tomcat Server Application. The tool incorporates several other open-source libraries and packages, a list of which is available along with the source code at request. 


\needspace{2em}\section{User Study}
We conducted a formative user study to gain insight into usability factors of the tool that may be associated with errors or confusion, and to identify opportunities for improvement via re-design or implementation of new functionality.

\needspace{2em}\subsection{Participants}
We adopted a snowball sampling technique starting with clinicians identified by our colleagues to recruit participants for the user study. We conducted a total of 4 (+1 pilot) studies lasting between 60 to 90 minutes. Our participants worked as both clinicians and clinical researchers and had at least an MD degree. All participants were experienced with both clinical text and the colonoscopy procedures. Their positions varied from research faculty members to physician scientists. Three out of the four participants had between 5-10 years of experience in that position. They had limited experience with machine learning algorithms with average self-reported proficiency being 5.0 on a scale of 1 to 10 (Individual ratings: 2, 5, 6, 7), where 1 is for \emph{``No knowledge at all"}, 5 -- \emph{``Some idea about the algorithms"}, and 10 being \emph{``Can read and understand current research"}. 

The pilot study helped us with some initial comments about the tool. This was done to identify any unnoticed bugs in our software prototype or any problems with our study protocol. Since we followed the same protocol in the pilot study as well and fixed only a couple of minor problems with the tool after it, we have also included its results with the rest of the studies. 

\needspace{2em}\subsection{Study Protocol}
We began with a pre-study survey to gauge background information about the participants and their expectations from the tool. We gave a short 15-minute walkthrough of the interface before handing over the control to them. During the study, the participants were asked review documents using the tool, and revise NLP predictions by providing feedback wherever required. We asked the participants to work and build models for only one of the variables -- \emph{biopsy}, indicating whether or not the report discussed a sample biopsy. Actual interactions with the tool lasted between 20-30 minutes for the 4 studies but was longer for the pilot. The participants worked with 280 documents for providing feedback for a model built against a set of 30 hand-annotated documents. We followed the ``think aloud" method to record their comments and reactions while using the tool. Sessions were conducted over web-conferencing software, which was also used to capture audio, screen content, and mouse interactions. At the end of the study, we asked users to complete the System Usability Scale~\cite{sus} and to answer some questions regarding their understanding of the tool. \hhcommentFixed{ I think this is ok now. 1. reference SUS explicitly. 2. you are generally doing a good job of avoiding passive tense, but too many sentences starting off with ``we did..'' gets to be a bit tiresome. try alternative phrasings.} 

\gtcomment{This is new. Needs review}
\begin{table}
\centering
  \small
    {
    \tabulinesep=1.2mm
    \begin{tabu}{|p{1.4cm}|p{0.05cm} p{5.95cm}|}
    \hline
    \multicolumn{1}{|c|}{\tabhead{Category}}    &  \multicolumn{2}{c|}{\tabhead{Recommendation}} \\ \hline
    \emph{Workflow} & 1. & Allow sorting (or filtering) of the documents in the grid based on the prediction probabilities. This would make it easier for the users to prioritize documents to review. \\ \cline{2-3}
                                     & 2. & Add a button to open the next-in-line document for review. The order may be decided either trivially based on ID number or by using an active learning approach. This would save the users to navigate through the grid when they don't have their own strategy for selecting documents for review.\\
    \hline
    \emph{WordTree} & 1. & Change the layout of the tool to show the WordTree view along with the document view. This would allow the user to quickly go through the full report text when the wordtree tree is unable to provide sufficient contextual information. \\ \cline{2-3}
                                     & 2. & Allow selection of multiple branches in the tree to give feedback on multiple paths in the tree at once.\\
    \hline
    \emph{Feedback} & 1. & Provide a feedback mechanism to specify that a text span does not indicate either of the classes. This would allow the user to remove non-informative but possibly misleading features in re-training. \\ \cline{2-3} \hline

    \emph{Re-Training} & 1. & Perform auto-retraining in the background when a sufficient number of feedback items have been provided by the user.\\ \cline{2-3}
    & 2. & Provide a built-in mechanism to validate and generate a performance report for the current model against a held-out test set.\\  \hline
    \end{tabu}
    }
    \caption{Summary of recommendations inferred from the user study for improving the interface design.}
  \label{tab:improvements}
\end{table}

\begin{table*}[ht]
  \centering
  \small
\begin{tabular}{R{0.39cm} | L{9.21cm}|L{1.05cm}|L{1.05cm}|L{1.05cm}|L{1.05cm}|L{1.05cm}}
    \tabhead{\#} & \tabhead{Question}                                                                                  & \tabhead{Strongly Disagree} & \tabhead{Disagree} & \tabhead{Neither} & \tabhead{Agree} & \tabhead{Strongly Agree} \\ \hline
    1.& I think that I'd like to use this system frequently                                   & 0                 & 1$^\dag$        & 0                          & 3     & 1              \\ \hline
    2.& I found this system unnecessarily complex                                                 & 0                 & 3        & 2                          & 0     & 0              \\ \hline
    3.& I thought the system was easy to use                                                      & 0                 & 0        & 0                          & 5     & 0              \\ \hline
    4.& I think that I'd need the support of a technical person to be able to use this system & 1                 & 2        & 0                          & 2     & 0              \\ \hline
    5.& I found the various functions in this system were well integrated                         & 0                 & 1        & 0                          & 4     & 0              \\ \hline
    6.& I thought there was too much inconsistency in this system                                 & 3                 & 1        & 1                          & 0     & 0              \\ \hline
    7.& I'd imagine that most people would learn to use this system very quickly              & 0                 & 0        & 1                          & 3     & 1              \\ \hline
    8.& I found the system very cumbersome to use                                                 & 1                 & 4        & 0                          & 0     & 0              \\ \hline
    9.& I felt very confident in using the system                                                 & 0                 & 0        & 1                          & 3     & 1              \\ \hline
    10.& I needed to learn a lot of things before I could get going with this system               & 0                 & 3        & 0                          & 2     & 0             
    \end{tabular}
  \caption{Frequency of responses used for SUS score \protect\cite{sus} calculation with
    5 participants. Average SUS score \hhcommentFixed{citation is in the previous sentence -- add citation}  was 70.5. a) We received poorer scores for questions  about the learnability of the tool (Q4, Q10) as we were only able to provide a limited walkthrough of the interface to our participants  due to time constraints. The participants acknowledged we could have done better if we were to spend more time on it. b) $^\dag$When asked whether they would use the system frequently, the participant remarked that he gave a low score because he did not see the methods being directly useful in his current work.}
  \label{tab:sus}
\end{table*}

\subsection{Results}
\subsubsection{System Usability scale}
We used the System Usability Scale consisting of 10-questions on a 5-point Likert scale to help get a global view of subjective assessments of usability. The average SUS score was 70.5 out of 100. Individual scores are provided in Table~\ref{tab:sus}.\hhcommentFixed{ok. I do have the proposed changes in this section but I have not summarized them; Should we do that in a table maybe? We do have space for it now --  summarize proposed design changes based on this feedback.}

\needspace{2em}\subsubsection{Semi-structured interviews after the study} 
We classified the collected observations from the think-aloud sessions and comments from the semi-structured interviews into four categories: 1) \emph{Workflow}: Comments and observations as the participants navigated through the documents for review, 2) \emph{WordTree}: While selecting search queries and browsing the wordtree, 3) \emph{Feedback}: While providing feedback to the learning system, and 4) \emph{Re-training:} Upon seeing changes after re-training. Some of these comments also include requests for new features by the participants which are also summarized in Table~\ref{tab:improvements}.

\begin{itemize}
\item \textbf{Workflow}: Participants used the grid view to select documents of interest and the document view to navigate through the text. They found this part of the workflow to be tedious. Some participants requested a ``Next" button that could be used to quickly move to a new document, instead of clicking on the cells in the grid. However, one of the participants also provided a contrasting view, expressing appreciation for the flexibility offered by the tool in selecting the documents for  labeling. They also made use of the color shades representing the prediction probability numbers to prioritize documents for inspection. They requested a sort feature in the grid view that could arrange the documents according to these probability scores as well.
\hhcommentFixed{detail is ok, but would be better to have count. 1. explain this detail. 2. Can you give specific counts of the number of users who made each requests?}  

\item \textbf{WordTree}: Perceptions about the wordtree were mixed. Some concerns regarding the wordtree appear to stem from the tabbed display that makes the user choose between the document view and the wordtree. While all participants found the wordtree to be a faster way to provide feedback, they felt that providing the feedback without being able to see the full document text at the same time might be error-prone. Although we were able to provide sentence long phrases in the tree and to show links to the full text of the relevant documents in the grid view, the participants were in favor of having a quicker way to access the complete reports. We have proposed a re-design to address this concern for future work. Our proposed redesign includes a provision for the user to make the WordTree view pop-out from its tab so that it can be used with the document and the grid views simultaneously. \hhcommentFixed{explain why it was risky - i.e., that the wordtree hid the document text.}

Participants discovered an unexpected use case for this view. In addition to giving feedback, the wordtree allowed users to verify the quality of their models. This was a consequence of the gradient colors in it, which showed how the presence of individual keywords affect the classification of documents. \hhcommentFixed{explain} By looking at how the gradient colors changed for the different keywords, the user could understand how well the model performed in predicting the values depending on the phrases contained in the document.

A common problem was that the users left the wordtree's search filter on  even after they were done using it. The tool would filter the documents in the grid as the users navigated through the wordtree but would forget to clear it for the next round of analysis. \hhcommentFixed{Is it okay now? -- explain} 

\item \textbf{Feedback}: Physicians indicated that they were accustomed to thinking in terms of rules suggesting a direct link between the feature and classification rather than the probabilistic associations used in our tool. As a result they were unsure at times about assigning a classification for a text span that serves as indicator in most but still not all of the cases. We address this problem in tool's design by encouraging models to be built iteratively. The user need not focus on building a completely accurate model at once but has an option to refine it for more specific cases in the future iterations. From the user study, however, we could not recommend any further design improvements that could make the users more comfortable with this workflow.  \hhcommentRevisit{i don't understand - what does this quote mean? clarify}


One missing feature pointed out by the users was the ability to select a phrase and say that it didn't contribute towards the classification of the documents, when it was being picked either as a true or a false feature by the learning system. Otherwise the participants found the tool's features very usable for sending feedback to the machine learning system.

\item \textbf{Re-Training}: We had suggested that participants could build as many models as they like, which led them to have doubts about the optimal frequency of retraining. Future work may use NLP metrics to automatically determine when to retrain. \hhcommentFixed{To be done in future work -- awkward - rephrase and clarify - you need to discuss how we might automatically decide when to retrain} The participants indicated that they were pleased to see the grid show changes in predictions after their feedback. Another suggestion was to provide a built-in option to test their model against a held-out hand annotated testing set.  
\end{itemize}

Overall we received very encouraging responses from the participants. Four out of five (including the pilot) expressed interest in having the tool made available for their own work right away. The remaining  participant was not involved in any research involving study clinical text. During the pre-study interview, we asked participants about their ideas on such a tool before showing our prototype. One of the participants who is actively working on related colonoscopy research requested features like a web-based interface for collaborating with people who are at geographically separated locations, flexibility in selecting documents to annotate, and a feedback mechanism for NLP. Our prototype tool was able to satisfy his needs in all of these aspects.

\needspace{2em}\section{Discussion and Future Work} \hhcommentFixed{summarize - develop the
tool and conducted preliminary evaluation.  -- Shouldn't this be a part of the conclusion}The initial feedback from the usability study provided both preliminary validation of the usability of the tool and guidance for improving the design of the tool. While we have not identified any major hurdle that would require a comprehensive re-design of any interface component, there are several extensions to the current set of features we believe might improve usability and will be promising in future work (Table~\ref{tab:improvements}). \hhcommentFixed{instead of these sentences, say that initial feedback from the usability study has provided guidance for designs that might improve usability.}

One of the aims of this project was to explore the feasibility of using interactive review as a means of lowering the training requirements. We hypothesize that the manual review supported by this tool will enable rapid convergence on highly accurate models even by starting with smaller training sets. Testing this out in a statistically compelling manner has been left for a future empirical evaluation study. \hhcommentFixed{discuss this as  a topic for further study - as below} This would involve observing efficiency measures such as overall time spent and accuracy measures like F-Measure etc. under different variations of the tool. We may control the tool's presentation capabilities, types of feedback allowed and the number of training documents as the independent variables during this study. Another promising future direction would be to evaluate the use of the tool by  several users in a collaborative work setting. \hhcommentRevisit{I have reordered things here -- too vague. just say that we will work towards an empirical evaluation.}

\needspace{2em}\section{Conclusion}

Despite the promising results shown by repeated studies involving NLP on clinical records, the benefits of NLP are all too often inaccessible to the clinicians and practitioners. Moreover, we have seen from previous studies that extracting structured insights from clinical text is hard. Although NLP techniques work well they have been put into limited use by the researchers in the field. In particular, without access to usable tools for clinicians that can make it easier for them to review and revise NLP findings, it is difficult apply these techniques. 

We have built a candidate tool to help address these problems. The interactive components of the tool along with novel visualization techniques support the entire interactive machine learning cycle with review, feedback and retraining steps. We conducted a user-study with prospective users as study participants to validate our design rationales. We also identified opportunities for improvement that will be addressed before we move forward with an empirical evaluation of the system.

\needspace{2em}\section{Acknowledgments}

We thank our user study participants. We would also like to thank Dr. Ateev Mehrotra for providing the colonoscopy reports dataset. This research was supported by NIH grant 5R01LM010964.

%
%
%
%
%
\balance

\bibliographystyle{acm-sigchi}
\vspace*{0.1mm}
\scriptsize
\renewcommand{\UrlFont}{\em\scriptsize}
\bibliography{sample}

\begin{thebibliography}{10}

\bibitem{iml-cuet}
Amershi, S., Lee, B., Kapoor, A., Mahajan, R., and Christian, B.
\newblock Cuet: Human-guided fast and accurate network alarm triage.
\newblock In {\em Proceedings of the SIGCHI Conference on Human Factors in
  Computing Systems}, CHI '11, ACM (New York, NY, USA, 2011), 157--166.

\bibitem{d3}
Bostock, M., Ogievetsky, V., and Heer, J.
\newblock D3: Data-driven documents.
\newblock {\em IEEE Trans. Visualization \& Comp. Graphics (Proc. InfoVis)\/}
  (2011).

\bibitem{sus}
Brooke, J.
\newblock {SUS:} a quick and dirty usability scale.
\newblock In {\em Usability evaluation in industry}, P.~W. Jordan,
  B.~Weerdmeester, A.~Thomas, and I.~L. Mclelland, Eds. Taylor and Francis,
  London, 1996.

\bibitem{barriers}
Chapman, W.~W., Nadkarni, P.~M., Hirschman, L., D'Avolio, L.~W., Savova, G.~K.,
  and Uzuner, O.
\newblock {{O}vercoming barriers to {N}{L}{P} for clinical text: the role of
  shared tasks and the need for additional creative solutions}.
\newblock {\em Journal of the American Medical Informatics Association 18}, 5
  (2011), 540--543.

\bibitem{iml-apolo}
Chau, D.~H., Kittur, A., Hong, J.~I., and Faloutsos, C.
\newblock Apolo: Making sense of large network data by combining rich user
  interaction and machine learning.
\newblock In {\em Proceedings of the SIGCHI Conference on Human Factors in
  Computing Systems}, CHI '11, ACM (New York, NY, USA, 2011), 167--176.

\bibitem{avolio}
D'Avolio, L.~W., Nguyen, T.~M., Goryachev, S., and Fiore, L.~D.
\newblock {{A}utomated concept-level information extraction to reduce the need
  for custom software and rules development}.
\newblock {\em Journal of the American Medical Informatics Association 18}, 5
  (2011), 607--613.

\bibitem{oldbook}
Dick, R.~S., Steen, E.~B., Detmer, D.~E., and {Institute of Medicine (U.S.).
  Committee on Improving the Patient Record}.
\newblock {\em The computer-based patient record: an essential technology for
  health care}.
\newblock National Academy Press, Washington, D.C, 1997.

\bibitem{iml-image}
Fails, J., and Olsen~Jr, D.
\newblock Interactive machine learning.
\newblock In {\em Proceedings of the 8th international conference on
  Intelligent user interfaces}, ACM (2003), 39--45.

\bibitem{iml-music}
Fiebrink, R., Cook, P.~R., and Trueman, D.
\newblock Human model evaluation in interactive supervised learning.
\newblock In {\em Proceedings of the SIGCHI Conference on Human Factors in
  Computing Systems}, CHI '11, ACM (New York, NY, USA, 2011), 147--156.

\bibitem{harkema}
Harkema, H., Chapman, W.~W., Saul, M., Dellon, E.~S., Schoen, R.~E., and
  Mehrotra, A.
\newblock Developing a natural language processing application for measuring
  the quality of colonoscopy procedures.
\newblock {\em JAMIA 18}, Supplement (2011), 150--156.

\bibitem{iml-heimerl}
Heimerl, F., Koch, S., Bosch, H., and Ertl, T.
\newblock Visual classifier training for text document retrieval.
\newblock {\em Visualization and Computer Graphics, IEEE Transactions on 18},
  12 (2012), 2839--2848.

\bibitem{adr-iyer}
Iyer, S.~V., Lependu, P., Harpaz, R., Bauer-Mehren, A., and Shah, N.~H.
\newblock {{L}earning signals of adverse drug-drug interactions from the
  unstructured text of electronic health records}.
\newblock {\em AMIA Summits Transl Sci Proc 2013\/} (2013), 83--87.

\bibitem{adr-medlee}
Melton, G.~B., and Hripcsak, G.
\newblock {Automated Detection of Adverse Events Using Natural Language
  Processing of Discharge Summaries}.
\newblock {\em Journal of the American Medical Informatics Association 12}, 4
  (2005), 448--457.

\bibitem{nlpintro}
Nadkarni, P.~M., Ohno-Machado, L., and Chapman, W.~W.
\newblock Natural language processing: an introduction.
\newblock {\em Journal of the American Medical Informatics Association 18}, 5
  (2011), 544--551.

\bibitem{narratives}
Nair, V., Kaduskar, M., Bhaskaran, P., Bhaumik, S., and Lee, H.
\newblock Preserving narratives in electronic health records.
\newblock In {\em Bioinformatics and Biomedicine (BIBM), 2011 IEEE
  International Conference on} (Nov 2011), 418--421.

\bibitem{knowtator}
Ogren, P.~V.
\newblock Knowtator: a prot\'{e}g\'{e} plug-in for annotated corpus
  construction.
\newblock In {\em Proceedings of the 2006 Conference of the North American
  Chapter of the Association for Computational Linguistics on Human Language
  Technology}, Association for Computational Linguistics (Morristown, NJ, USA,
  2006), 273--275.

\bibitem{adr-ramirez}
Ramirez, E., Carcas, A.~J., Borobia, A.~M., Lei, S.~H., Pinana, E., Fudio, S.,
  and Frias, J.
\newblock {{A} pharmacovigilance program from laboratory signals for the
  detection and reporting of serious adverse drug reactions in hospitalized
  patients}.
\newblock {\em Clin. Pharmacol. Ther. 87}, 1 (Jan 2010), 74--86.

\bibitem{ctakes}
Savova, G.~K., Masanz, J.~J., Ogren, P.~V., Zheng, J., Sohn, S.,
  Kipper-Schuler, K.~C., and Chute, C.~G.
\newblock {{M}ayo clinical {T}ext {A}nalysis and {K}nowledge {E}xtraction
  {S}ystem (c{T}{A}{K}{E}{S}): architecture, component evaluation and
  applications}.
\newblock {\em Journal of the American Medical Informatics Association 17}, 5
  (2010), 507--513.

\bibitem{silverasm-github}
Shrikumar, A.
\newblock A quick-and-dirty wordtree visualization in {JS}, {HTML} and {SVG}.
\newblock \url {https://github.com/silverasm/wordtree}, 2011.

\bibitem{jigsaw}
Stasko, J., G\"{o}rg, C., and Liu, Z.
\newblock Jigsaw: Supporting investigative analysis through interactive
  visualization.
\newblock {\em Information Visualization 7}, 2 (Apr. 2008), 118--132.

\bibitem{WordTree}
Wattenberg, M., and Viegas, F.~B.
\newblock The word tree, an interactive visual concordance.
\newblock {\em IEEE Transactions on Visualization and Computer Graphics 14}, 6
  (2008), 1221--1228.

\bibitem{tiara}
Wei, F., Liu, S., Song, Y., Pan, S., Zhou, M.~X., Qian, W., Shi, L., Tan, L.,
  and Zhang, Q.
\newblock Tiara: A visual exploratory text analytic system.
\newblock In {\em Proceedings of the 16th ACM SIGKDD International Conference
  on Knowledge Discovery and Data Mining}, KDD '10, ACM (New York, NY, USA,
  2010), 153--162.

\bibitem{yeye}
Ye, Y., Tsui, F.~R., Wagner, M., Espino, J.~U., and Li, Q.
\newblock Influenza detection from emergency department reports using natural
  language processing and bayesian network classifiers.
\newblock {\em Journal of the American Medical Informatics Association\/}
  (2014).

\end{thebibliography}
\end{document}